\documentclass[a4paper,11pt]{article}
\usepackage{pos}
\usepackage{subfigure}

\title{Dark sector searches at BESIII}
\author*[a]{Zhi-Jun Li}
\author[a]{Zheng-Yun You}
\affiliation[a]{School of Physics, Sun Yat-sen University, Guangzhou 510275, China}
\emailAdd{lizhj37@mail2.sysu.edu.cn}
\emailAdd{youzhy5@mail.sysu.edu.cn}

\abstract{Large samples of charmonia, charm mesons, hyperons and light mesons collected by BESIII offer sensitive tests of dark sector interactions with charm and light quarks. This contribution reviews recent searches for sub-GeV dark matter in $\eta\to\pi^0+\mathrm{invisible}$ and $J/\psi\to\phi+\mathrm{invisible}$, massless dark photons in $D^0\to\omega\gamma'$ and $D^0\to\gamma\gamma'$, invisible $K^0_S$ decays, dark baryons in $\Xi^-\to\pi^-\chi$, and light vector bosons in $\chi_{cJ}\to J/\psi V$, $V\to e^+e^-$. Particular attention is given to the $\eta$ search, which constrains scalar-mediated dark matter and tests thermal relic benchmarks.}

\FullConference{XVI International Conference on Beauty, Charm, Hyperons in Hadronic Interactions (BEACH 2026)\\
7-12 June 2026\\
Firenze, Italy\\}

\begin{document}
\maketitle

\section{Introduction}
Despite its success, the Standard Model (SM) leaves the nature of dark matter (DM), the strong CP problem, the fermion mass hierarchy and the matter--antimatter asymmetry unresolved. These open questions motivate searches for new physics (NP), including a dark sector with light particles accessible at high-intensity $e^+e^-$ experiments. BESIII~\cite{BESIII:2009fln}, operating at BEPCII~\cite{Yu:2016cof} in the $\tau$--charm region, provides a suitable setting for such studies. Its datasets include approximately $10^{10}$ $J/\psi$ and $2.7\times10^9$ $\psi(2S)$ events, $20~\mathrm{fb}^{-1}$ at $3.773~\mathrm{GeV}$, and more than $20~\mathrm{fb}^{-1}$ above $4.0~\mathrm{GeV}$~\cite{Liao:2025lth}. Together with the hyperons and light mesons produced in charmonium decays, these samples and improved analysis tools~\cite{Li:2024pox,Song:2025pnt} support a broad programme of dark sector searches.

\section{\texorpdfstring{Search for sub-GeV DM in $\eta\to\pi^0+\rm{invisible}$}{Search for sub-GeV DM in eta to pi0 + invisible}}
Astrophysical evidence requires a DM component that is absent from the SM. Direct searches based on elastic scattering from nuclei strongly constrain GeV-scale DM, but lose sensitivity at lower masses because the nuclear recoil energy often falls below the detection threshold. Meson decays at BESIII provide a complementary route to the sub-GeV region.

If a dark boson mediates both DM production and DM--nucleon scattering, $\eta\to\pi^0\chi\bar\chi$ and $\chi N\to\chi N$ probe the same interaction, allowing collider limits to be compared with direct-detection results. Here $\chi$ denotes a DM particle and $N$ a nucleon. For a fixed coupling, the flavour-neutral character and narrow width of the $\eta$ favour a large branching fraction (BF) relative to other meson decays. We consider a scalar mediator $S$. Although Higgs-like scalar couplings suppress interactions with light quarks, a flavour-specific scalar can couple appreciably to them~\cite{Batell:2018fqo}, making light-meson decays particularly relevant.

The search uses $(10\,087\pm44)\times10^6$ $J/\psi$ events. In $J/\psi\to K^+K^-\eta$, the recoil against the charged-kaon pair tags the $\eta$ independently of its decay, yielding $(2185.3\pm3.4)\times10^3$ inclusive $\eta$ candidates. An additional reconstructed $\pi^0$ selects $\eta\to\pi^0S$, with the invisible $S\to\chi\bar\chi$ system inferred from the recoil against $K^+K^-\pi^0$. No significant excess is found for $m_S=0$--$400~\mathrm{MeV}$. The resulting upper limits (ULs) on $\mathcal B(\eta\to\pi^0S\to\pi^0\chi\bar\chi)$ range from $1.8\times10^{-5}$ to $5.5\times10^{-5}$ at the $90\%$ confidence level (CL)~\cite{BESIII:2026qsu}.

\begin{figure}[htbp]
\centering
\subfigure[]{\includegraphics[width=0.49\textwidth]{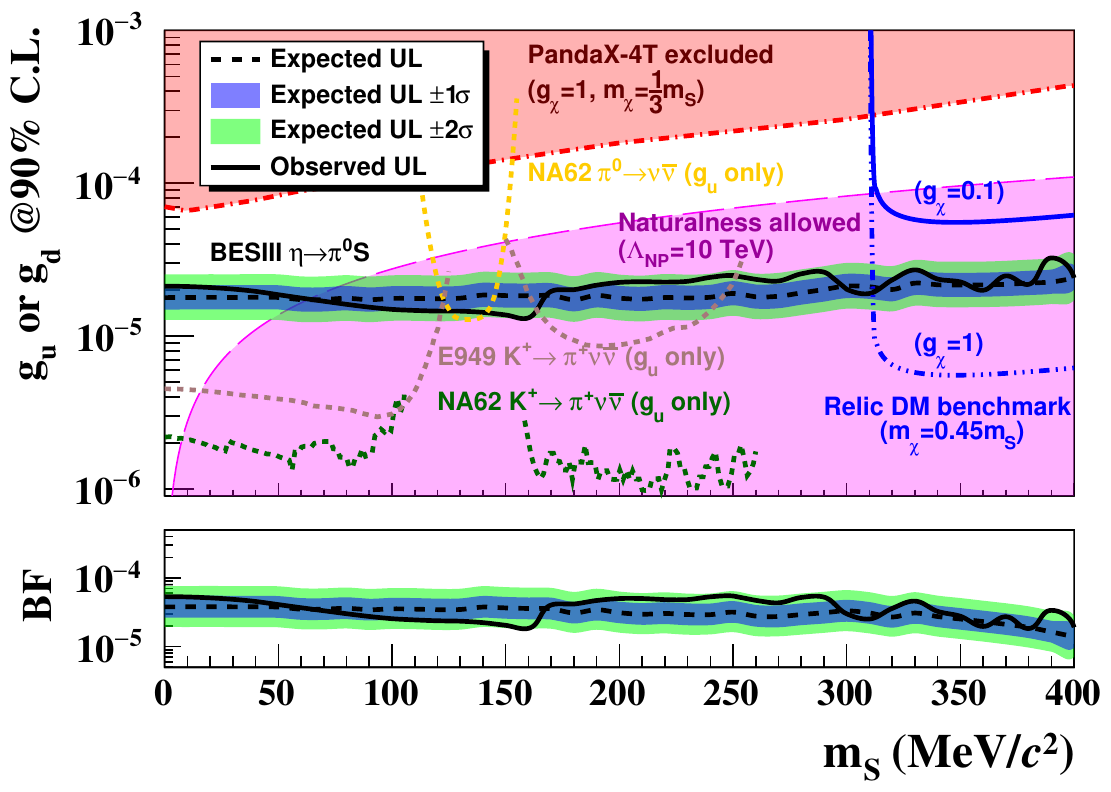}}
\subfigure[]{\includegraphics[width=0.49\textwidth]{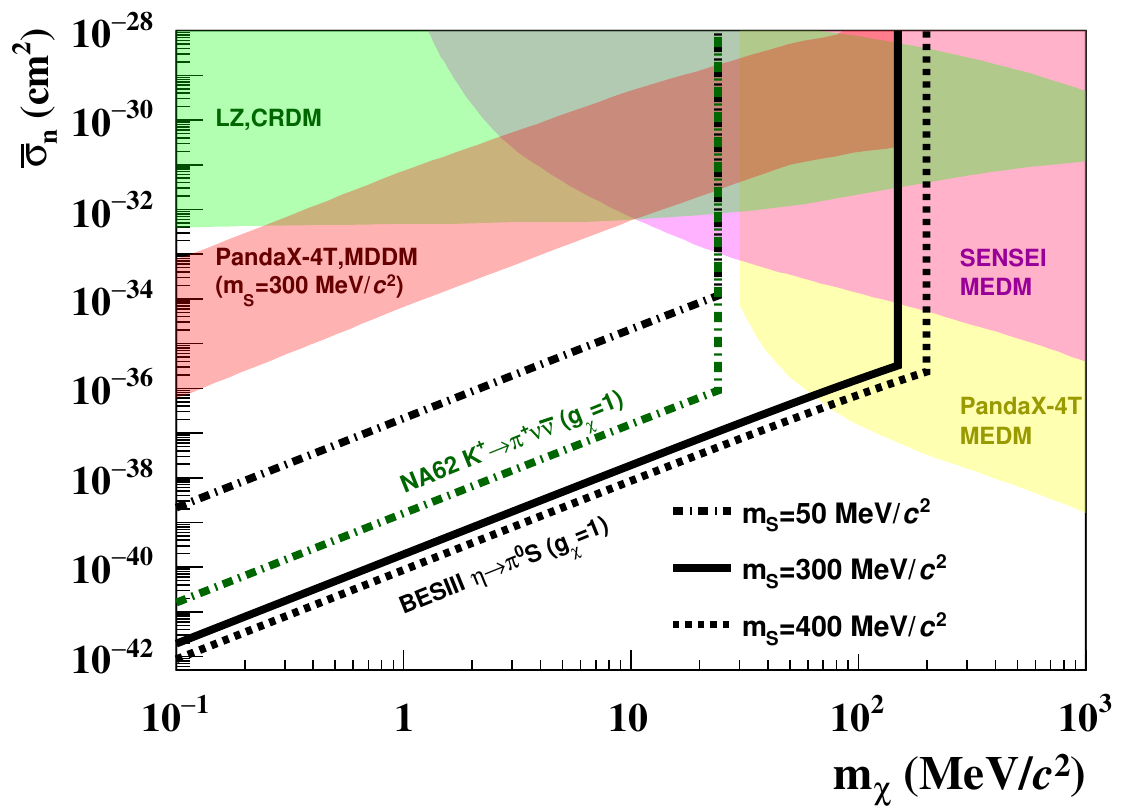}}
\caption{(a) $90\%$ CL limits on $g_u$ or $g_d$ (upper panel) and $\mathcal B(\eta\to\pi^0S)$ (lower panel). The purple region satisfies the naturalness bound for $\Lambda_{\mathrm{NP}}=10~\mathrm{TeV}$; the blue curves are thermal relic benchmarks for $g_\chi=0.1$ and $1$. The kaon limits apply only to $g_u$. (b) Limits on the reference DM--nucleon cross section $\bar\sigma_n$, compared with scattering experiments and NA62, for the assumptions indicated.}
\label{fig:eta2pi0}
\end{figure}

The flavour-specific interaction is described by~\cite{Batell:2018fqo}
\begin{equation}
\mathcal L\supset-g_\chi S\bar\chi_L\chi_R-g_uS\bar u_Lu_R+\mathrm{h.c.},
\label{eq:scalar}
\end{equation}
where $g_\chi$ and $g_u$ couple $S$ to DM and up quarks, respectively, and $L,R$ denote chirality. Replacing the up-quark term by its down-quark counterpart defines $g_d$. As shown in Figure~\ref{fig:eta2pi0}(a), the $90\%$ CL coupling limits are $(1.3$--$3.2)\times10^{-5}$, improving on the scalar-mediated PandaX-4T result~\cite{PandaX:2023tfq} by factors of $3.3$--$18.3$. Kaon-sector searches~\cite{NA62:2025upx,BNL-E949:2009dza,NA62:2020pwi} constrain only $g_u$, whereas the $\eta$ channel also probes $g_d$.

The naturalness region in Figure~\ref{fig:eta2pi0}(a) follows the effective field theory (EFT) bound $g_u\leq(16\pi^2/\sqrt2)m_S\nu/\Lambda_{\mathrm{NP}}^2$, with the Higgs vacuum expectation value $\nu=246.2~\mathrm{GeV}$~\cite{Batell:2018fqo}. Choosing $\Lambda_{\mathrm{NP}}=10~\mathrm{TeV}$ illustrates how low-energy $\eta$ decays test parameter space associated with a much higher NP scale. The thermal relic curves assume $m_\chi=0.45m_S$ and freeze-out through $\chi\bar\chi\to S\to\pi\pi$, reproducing $\Omega_{\mathrm{DM}}h^2=0.12$. The search excludes part of this benchmark for $g_\chi=0.1$; probing the $g_\chi=1$ case requires more data.

The same scalar-exchange model translates the coupling bounds into limits on the reference DM--nucleon scattering cross section $\bar\sigma_n$, shown in Figure~\ref{fig:eta2pi0}(b) for $g_\chi=1$. Since $\bar\sigma_n\propto g_\chi^2$, smaller assumed DM couplings give lower cross-section limits. In the relevant sub-GeV parameter space, BESIII improves on previous scattering constraints by approximately five orders of magnitude, demonstrating the reach of this meson-decay approach.

\section{\texorpdfstring{Search for sub-GeV DM in $J/\psi\to\phi+\mathrm{invisible}$}{Search for sub-GeV DM in J/psi to phi + invisible}}
An independent search uses $(8774.0\pm39.4)\times10^6$ $J/\psi$ events and reconstructs $\phi\to K^+K^-$. The recoil mass of the kaon pair identifies the invisible system; requiring it to lie below $2m_{K^0_L}$ reduces backgrounds. With no significant signal, the $90\%$ CL bound is $\mathcal B(J/\psi\to\phi+\mathrm{invisible})<7.0\times10^{-8}$. For a two-body decay to a single invisible particle, mass-dependent limits range from $4\times10^{-9}$ to $4\times10^{-8}$. The same analysis gives $\mathcal B(\eta\to\mathrm{invisible})<2.4\times10^{-5}$, improving the earlier bound by more than a factor of four~\cite{BESIII:2025cyj}. Interpreting these BF limits as coupling constraints requires further model-specific calculations.

\section{\texorpdfstring{Search for massless dark photons in $D^0\to\omega\gamma'$ and $D^0\to\gamma\gamma'$}{Search for massless dark photons in D0 decays}}
A massless dark photon $\gamma'$ arises from an unbroken additional Abelian gauge symmetry. It can mediate a long-range dark force; related models address flavour, fermion masses and vacuum stability~\cite{Fabbrichesi:2020wbt}. Interactions with SM matter can be described by dimension-six operators~\cite{Dobrescu:2004wz}:
\begin{equation}
\begin{aligned}
\mathcal L_{\mathrm{NP}}=\frac{1}{\Lambda_{\mathrm{NP}}^2}\bigl(&C^u_{jk}\bar q_j\sigma^{\mu\nu}u_k\widetilde H+C^d_{jk}\bar q_j\sigma^{\mu\nu}d_kH+C^l_{jk}\bar l_j\sigma^{\mu\nu}e_kH+\mathrm{h.c.}\bigr)F'_{\mu\nu}.
\end{aligned}
\label{eq:dimension-six}
\end{equation}
Here $C_{jk}$ are dimensionless, $q,l$ are fermion doublets, $u,d,e$ are singlets, and $H,F'_{\mu\nu}$ denote the Higgs doublet and dark-photon field strength. Flavour-changing terms permit $c\to u\gamma'$.

\begin{figure}[htbp]
\centering
\subfigure[]{\includegraphics[width=0.32\textwidth]{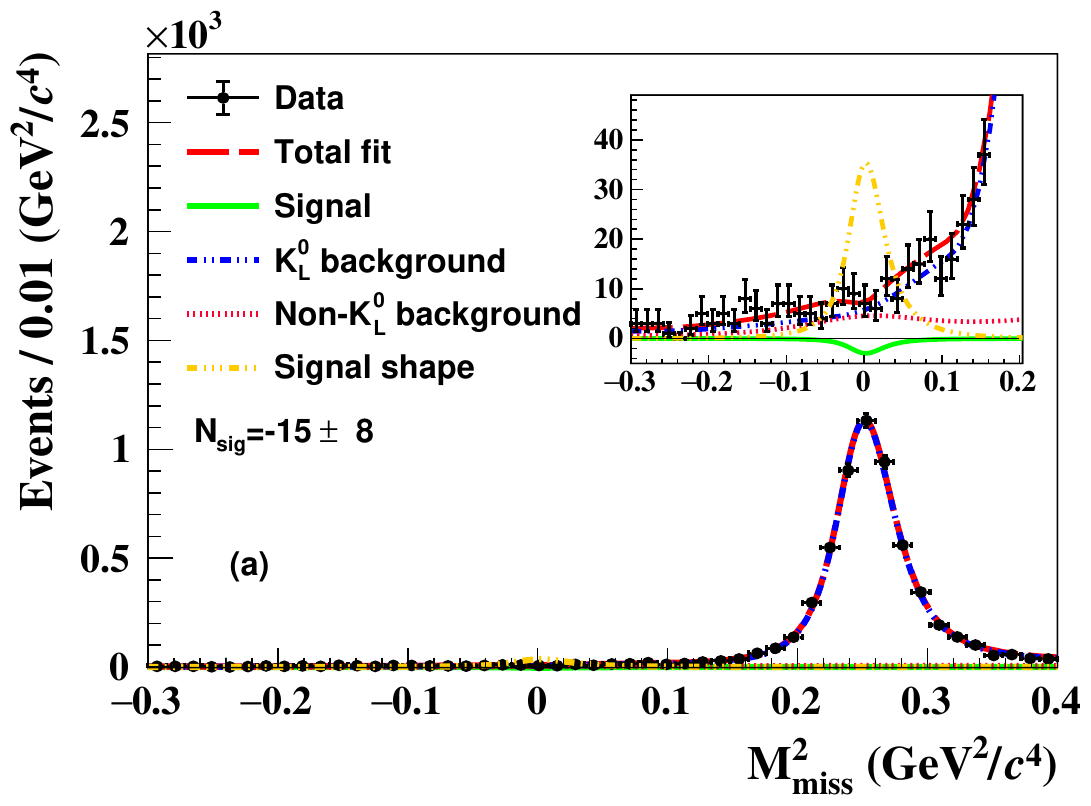}}
\subfigure[]{\includegraphics[width=0.32\textwidth]{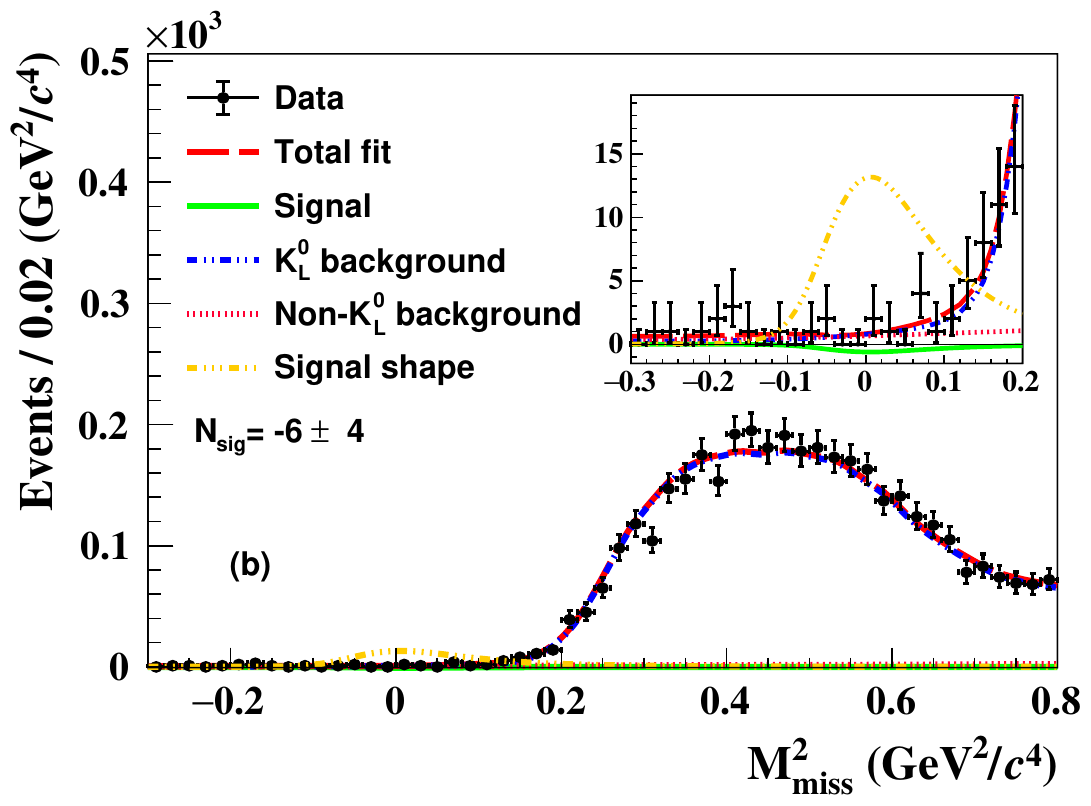}}
\subfigure[]{\includegraphics[width=0.32\textwidth]{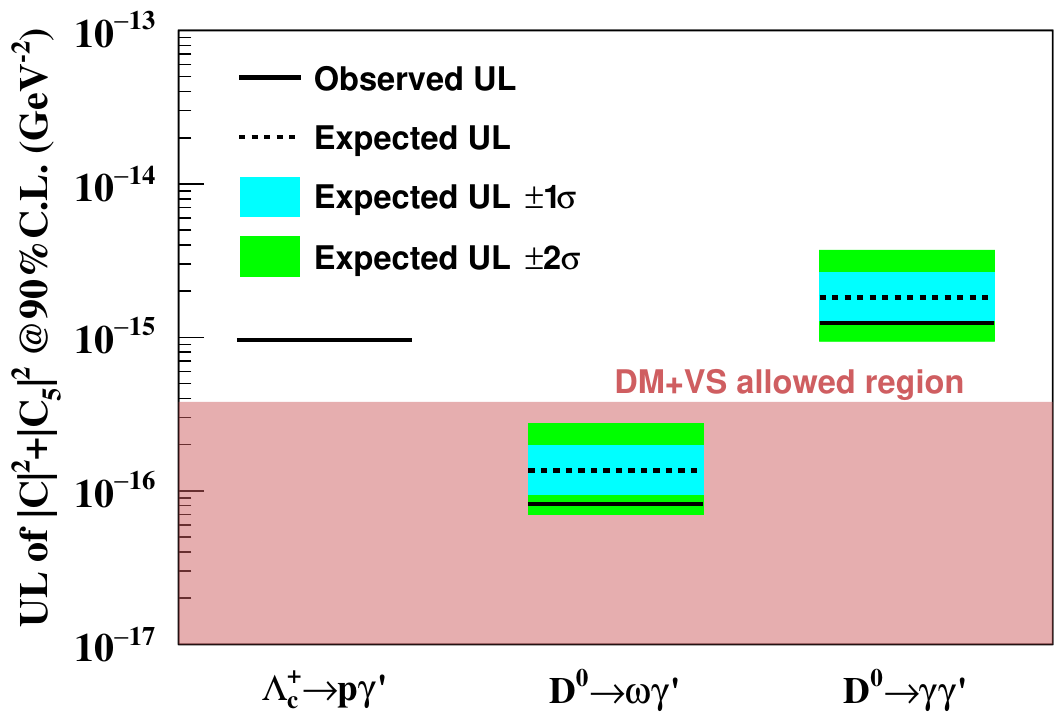}}
\caption{Missing-mass-squared distributions for (a) $D^0\to\omega\gamma'$ and (b) $D^0\to\gamma\gamma'$, with data, fitted components and illustrative signal shapes. (c) Limits on $|\mathcal C|^2+|\mathcal C_5|^2$; the red band denotes the region allowed by DM and vacuum stability (VS).}
\label{fig:Gp}
\end{figure}

Using $7.9~\mathrm{fb}^{-1}$ at $\sqrt s=3.773~\mathrm{GeV}$, BESIII tags one member of a $D^0\bar D^0$ pair in hadronic modes and searches the recoiling side for $\omega\gamma'$ or $\gamma\gamma'$. The missing-mass-squared distributions in Figure~\ref{fig:Gp}(a,b) show no significant signal. The respective $90\%$ CL BF limits are $1.1\times10^{-5}$ and $2.0\times10^{-6}$~\cite{BESIII:2024rkp}.

Both BFs scale with $Q=|\mathcal C|^2+|\mathcal C_5|^2$~\cite{Su:2020yze}: $\mathcal B(D^0\to V\gamma')=\tau_D f_{DV}^2(m_D^2-m_V^2)^3Q/(2\pi m_D^3)$ and $\mathcal B(D^0\to\gamma\gamma')=(\alpha_e/2)\tau_D f_{D\gamma}^2m_D^3Q$, where $f$, $\tau_D$ and $\alpha_e$ are the form factors, lifetime and fine-structure constant. Here $\mathcal C=\Lambda_{\mathrm{NP}}^{-2}(C^u_{12}+C^{u*}_{12})\nu/\sqrt8$ and $\mathcal C_5=\Lambda_{\mathrm{NP}}^{-2}(C^u_{12}-C^{u*}_{12})\nu/\sqrt8$. The $\omega\gamma'$ bound, $Q<8.2\times10^{-17}~\mathrm{GeV}^{-2}$, improves on $\Lambda_c^+\to p\gamma'$ by more than an order of magnitude and first reaches the DM- and VS-allowed region [Figure~\ref{fig:Gp}(c)]. The electromagnetic factor makes the $\gamma\gamma'$ coupling constraint weaker despite its smaller BF limit.

\section{\texorpdfstring{Search for the invisible decay of $K^0_S$}{Search for the invisible decay of KS}}
Invisible $K^0_S$ decays are suppressed by loops and helicity in the SM, but DM final states or ordinary--mirror oscillations could give BFs near $10^{-6}$. Constraining these modes also supports CPT tests through the Bell--Steinberger relation, which involves all neutral-kaon decay amplitudes and usually neglects invisible channels~\cite{Gninenko:2014sxa}.

\begin{figure}[htbp]
\centering
\subfigure[]{\includegraphics[width=0.60\textwidth]{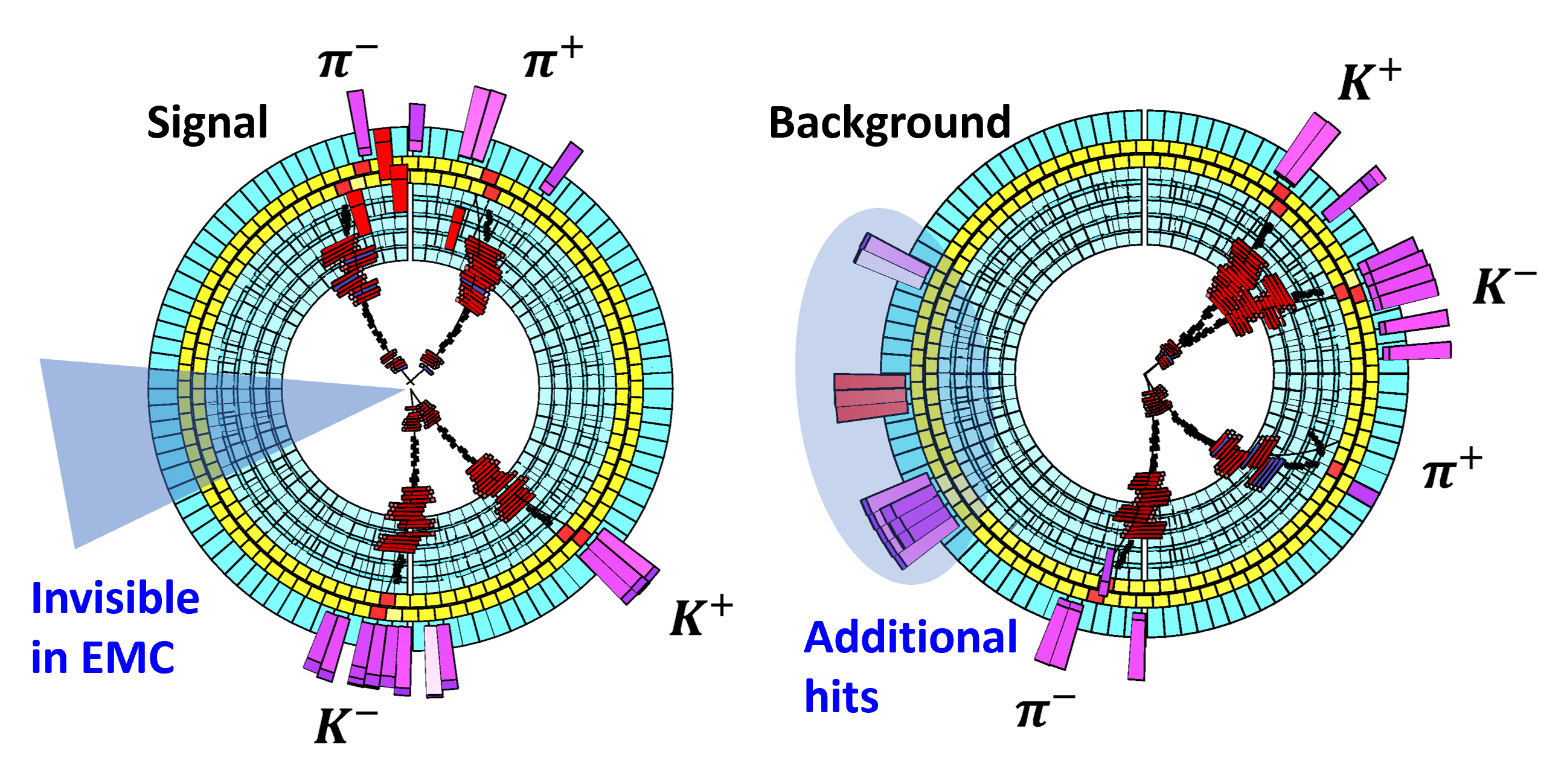}}
\subfigure[]{\includegraphics[width=0.37\textwidth]{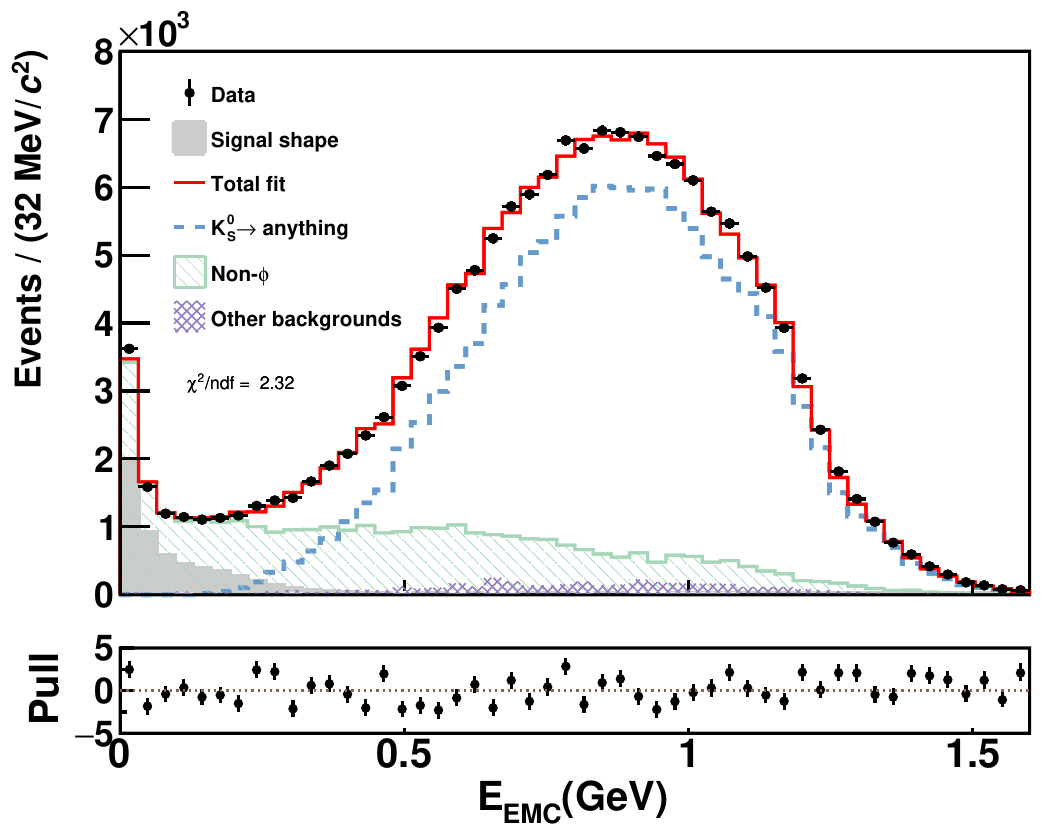}}
\caption{(a) Simulated invisible-$K^0_S$ signal (left) and background (right) event displays; the light-blue barrel is the EMC and purple blocks mark energy deposits. (b) Unused EMC energy in the selected data, with fitted components and the signal shape.}
\label{fig:Ks}
\end{figure}

The first direct search selects $J/\psi\to\phi K^0_SK^0_S$ from $10^{10}$ $J/\psi$ events. C parity suppresses $\phi K^0_SK^0_L$ backgrounds. Reconstructed $\phi\to K^+K^-$ and $K^0_S\to\pi^+\pi^-$ decays tag the second kaon, whose invisible decay leaves no extra electromagnetic calorimeter (EMC) activity [Figure~\ref{fig:Ks}(a)]. The EMC-energy fit [Figure~\ref{fig:Ks}(b)] includes a near-zero non-$\phi$ $K^+K^-K^0_SK^0_L$ contribution constrained by $\phi$ sidebands. The null result gives $\mathcal B(K^0_S\to\mathrm{invisible})<8.4\times10^{-4}$ at $90\%$ CL~\cite{BESIII:2025kjj}, above the NP predictions.

\section{\texorpdfstring{Search for dark baryons in $\Xi^-\to\pi^-\chi$}{Search for dark baryons in Xi decays}}
Dark baryons could connect the cosmic densities $\rho_{\mathrm{DM}}\simeq5.4\rho_{\mathrm{baryon}}$ and explain the neutron beam--storage lifetime difference through a roughly $1\%$ exotic decay fraction. CP-violating $B$ decays involving dark baryons can produce DM and the baryon asymmetry in $B$-mesogenesis~\cite{Alonso-Alvarez:2021oaj}; hyperons provide complementary tests with light quarks.

\begin{figure}[htbp]
\centering
\subfigure[]{\includegraphics[width=0.30\textwidth]{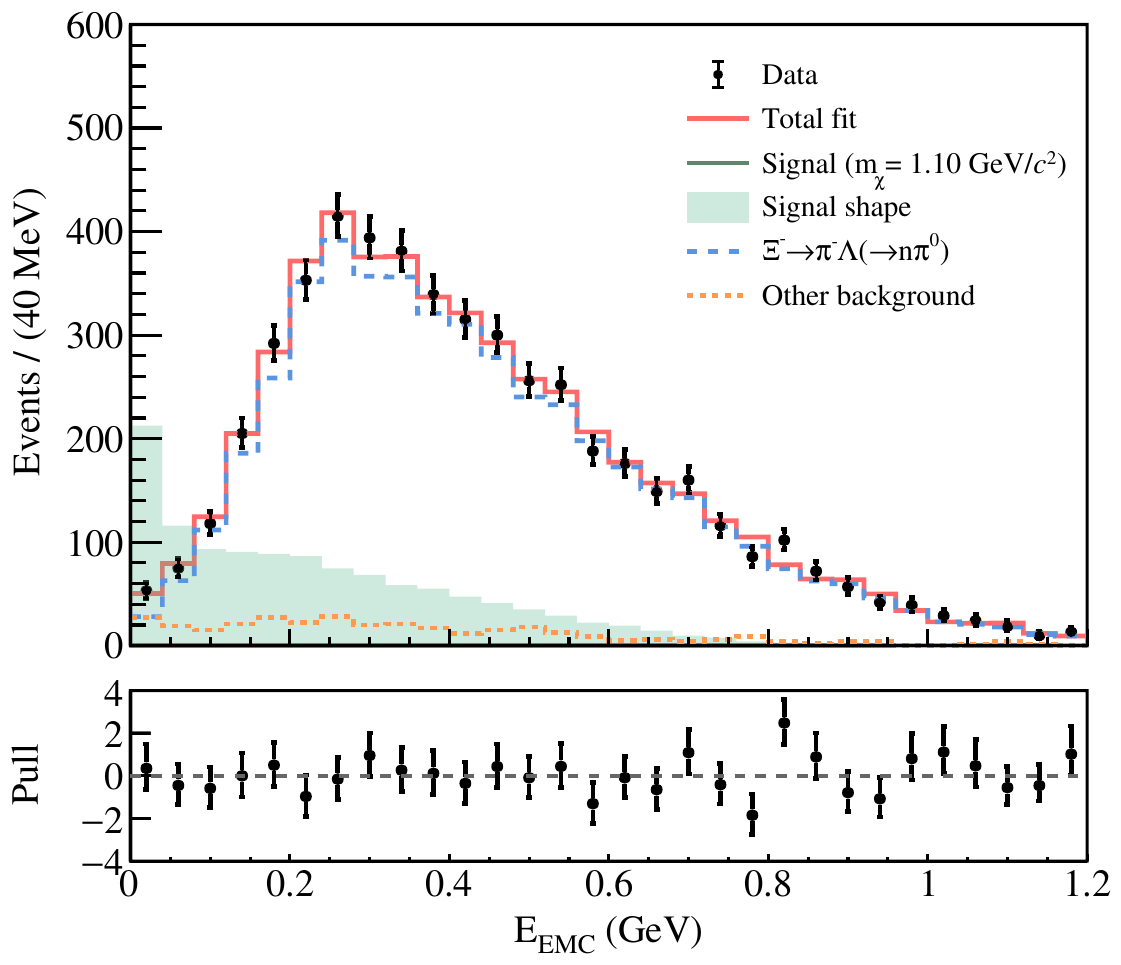}}
\subfigure[]{\includegraphics[width=0.34\textwidth]{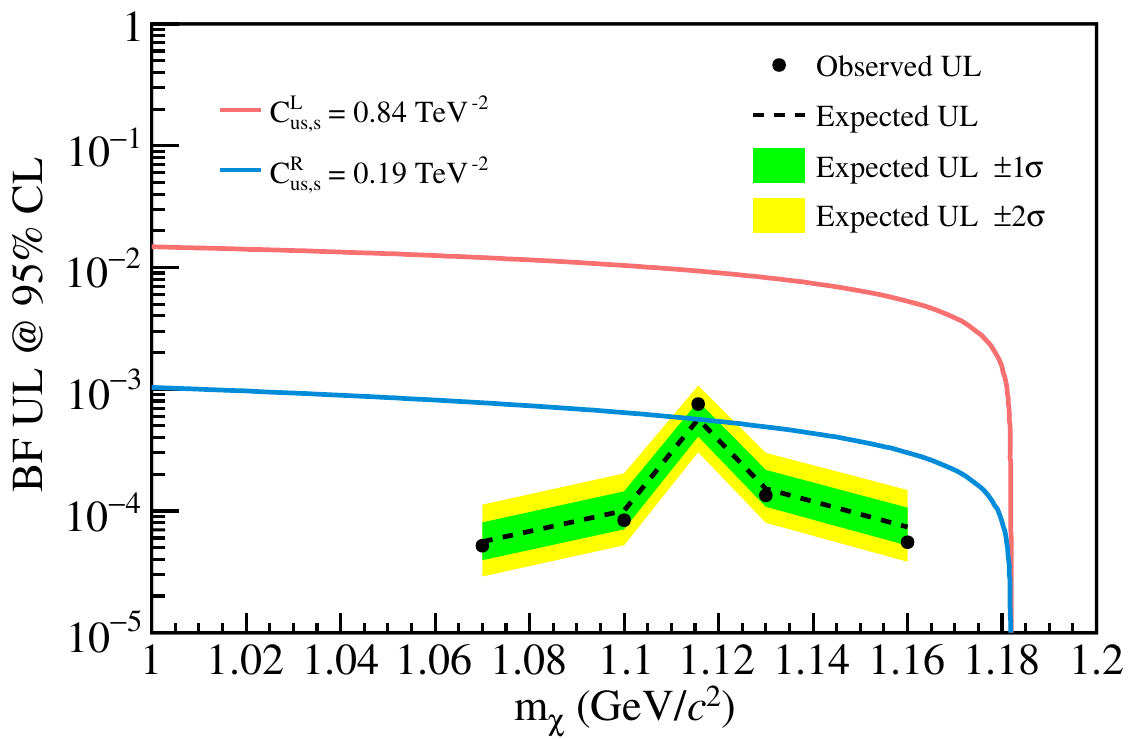}}
\subfigure[]{\includegraphics[width=0.32\textwidth]{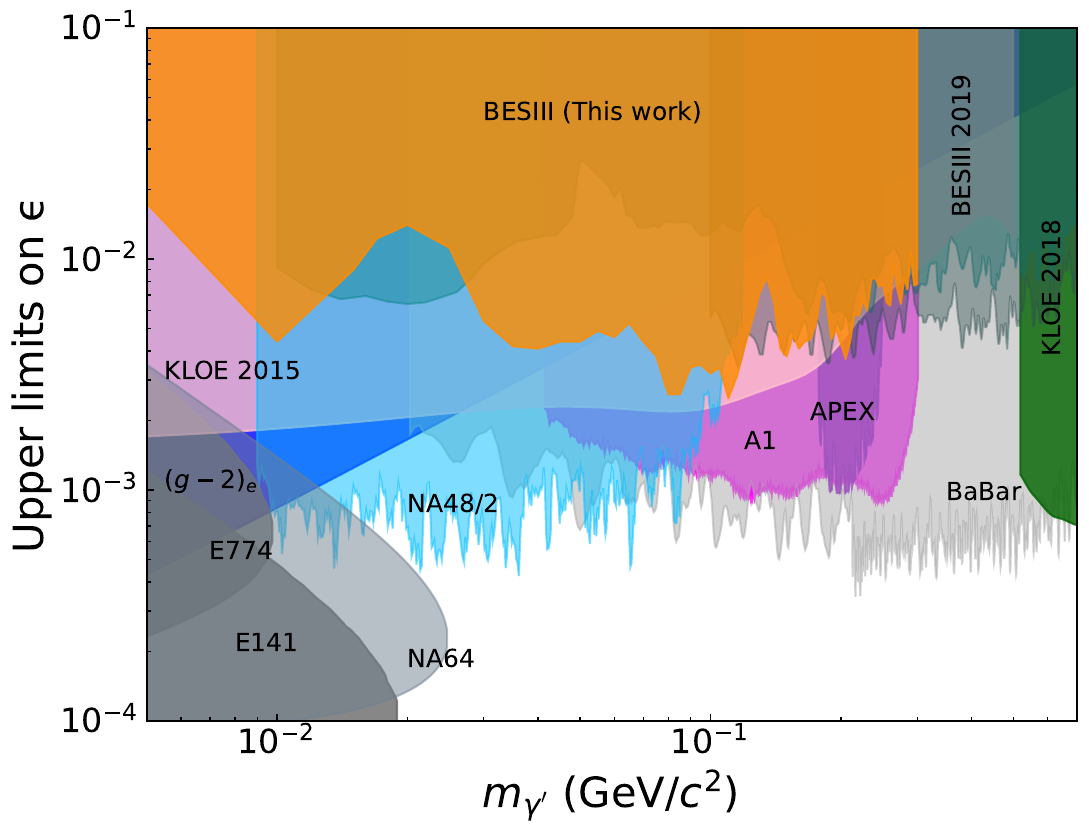}}
\caption{(a) Unused EMC energy for $\Xi^-\to\pi^-\chi$ at $m_\chi=1.10~\mathrm{GeV}$. (b) Observed and expected $95\%$ CL BF limits, compared with the recast bounds of Ref.~\cite{Alonso-Alvarez:2021oaj}. (c) $90\%$ CL limits on the dark-photon mixing parameter $\epsilon$ from charmonium transitions, compared with other searches~\cite{BESIII:chic-vector}.}
\label{fig:Xi}
\end{figure}

The first BESIII search for $\Xi^-\to\pi^-\chi$ uses about $10^7$ $\Xi^-\bar\Xi^+$ pairs from $(10\,084\pm44)\times10^6$ $J/\psi$ decays~\cite{BESIII:2025sfl}. A reconstructed $\bar\Xi^+\to\pi^+\bar\Lambda$, $\bar\Lambda\to\bar p\pi^+$ tags the $\Xi^-$. Kinematic fits constrain the missing mass to $m_\chi=1.07$, $1.10$, $m_\Lambda$, $1.13$ or $1.16~\mathrm{GeV}$. Fits to the unused EMC energy reveal no signal at any tested mass; Figure~\ref{fig:Xi}(a) shows the $1.10~\mathrm{GeV}$ case. The $95\%$ CL BF limits in Figure~\ref{fig:Xi}(b) improve on the LHC bounds recast in Ref.~\cite{Alonso-Alvarez:2021oaj} within the dark-baryon model.

\section{\texorpdfstring{Search for light vector bosons in $\chi_{cJ}\to J/\psi V$, $V\to e^+e^-$}{Search for light vector bosons in charmonium transitions}}
The visible channel $\chi_{cJ}\to J/\psi V$ tests the interaction $\mathcal L\supset\sum_f eq_f\epsilon_fV^\mu\bar\psi_f\gamma_\mu\psi_f$, where $eq_f$ is the fermion charge and $\epsilon_f$ its relative coupling. BESIII produces $\chi_{cJ}$ through $\psi(2S)\to\gamma\chi_{cJ}$ in $(2712.4\pm14.3)\times10^6$ $\psi(2S)$ events. No narrow $e^+e^-$ peak is found~\cite{BESIII:chic-vector}. For dark photons, $\epsilon_f=\epsilon$ universally: the $90\%$ CL bounds are $(2.5$--$17.5)\times10^{-3}$ over $5$--$300~\mathrm{MeV}$ [Figure~\ref{fig:Xi}(c)]. These limits are weaker than other dark-photon bounds, but the channel directly probes charm-quark couplings in non-universal models.

\section{Summary}
BESIII sets first or improved bounds on invisible meson decays, dark baryons and light vector bosons. In particular, $\eta\to\pi^0+\mathrm{invisible}$ strengthens scalar-mediated sub-GeV DM constraints and tests thermal relic benchmarks. The accumulated $\tau$--charm samples enable further searches and improved interpretations.

\section*{Acknowledgements}
This work is supported in part by the NSFC (Grant No. 125B2107) and the National Key R\&D Program of China (Grant No. 2023YFA1606000).

\end{document}